\documentclass{IEEEtran4PSCC}
\ifCLASSINFOpdf
  \usepackage[pdftex]{graphicx}
\else
  \usepackage[dvips]{graphicx}
\fi
\usepackage[cmex10]{amsmath}
\usepackage{xcolor}
\usepackage{amsmath,amssymb}

\global\long\def\argmin{\operatornamewithlimits{argmin}}

\renewcommand{\IEEEQED}{\IEEEQEDopen}

\DeclareMathOperator{\Tr}{Tr}
\newtheorem{thm}{Theorem}
\newtheorem{prop}[thm]{Proposition}
\newtheorem{cor}[thm]{Corollary}

\begin{document}
%
% paper title
% Titles are generally capitalized except for words such as a, an, and, as,
% at, but, by, for, in, nor, of, on, or, the, to and up, which are usually
% not capitalized unless they are the first or last word of the title.
% Linebreaks \\ can be used within to get better formatting as desired.
% Do not put math or special symbols in the title.
\title{Online Learning of Power Transmission Dynamics}

% % To specify the authors when (number of affiliations <= 2)
% \author{
% \IEEEauthorblockN{Andrey Y. Lokhov, Marc Vuffray, Deepjyoti Deka, Michael Chertkov
% }
% \IEEEauthorblockA{Theoretical Division, Los Alamos National Laboratory, Los Alamos, NM 87544 United States\\
% \{lokhov, vuffray, deepjyoti, chertkov\}@lanl.gov}
% }

% To specify the authors when (number of affiliations > 2)
\author{\IEEEauthorblockN{Andrey Y. Lokhov\IEEEauthorrefmark{1}\IEEEauthorrefmark{2},
Marc Vuffray\IEEEauthorrefmark{1},
Dmitry Shemetov\IEEEauthorrefmark{3}, 
Deepjyoti Deka\IEEEauthorrefmark{1} and
Michael Chertkov\IEEEauthorrefmark{1}\IEEEauthorrefmark{2}\IEEEauthorrefmark{4}}
% Template-style formatting
% \IEEEauthorblockA{\IEEEauthorrefmark{1} Theoretical Division\\
% Los Alamos National Laboratory,
% Los Alamos, NM 87544\\ \{lokhov,vuffray,deepjyoti,chertkov\}@lanl.gov}
\IEEEauthorblockA{\IEEEauthorrefmark{1} Theoretical Division,
Los Alamos National Laboratory,
Los Alamos, NM 87544}
%\{lokhov,vuffray,deepjyoti,chertkov\}@lanl.gov
\IEEEauthorblockA{\IEEEauthorrefmark{2} Center for Nonlinear Studies,
Los Alamos National Laboratory,
Los Alamos, NM 87544}
\IEEEauthorblockA{\IEEEauthorrefmark{3} Complexity Sciences Center,
University of California at Davis,
Davis, CA 95616}
%, dshemetov@ucdavis.edu
\IEEEauthorblockA{\IEEEauthorrefmark{4} Skolkovo Institute of Science and Technology,
143026 Moscow, Russia}
}

% make the title area
\maketitle

% As a general rule, do not put math, special symbols or citations
% in the abstract
\begin{abstract}
We consider the problem of reconstructing the dynamic state matrix of transmission power grids from time-stamped PMU measurements in the regime of ambient fluctuations. Using a maximum likelihood based approach, we construct a family of convex estimators that adapt to the structure of the problem depending on the available prior information. The proposed method is fully data-driven and does not assume any knowledge of system parameters. It can be implemented in near real-time and requires a small amount of data. Our learning algorithms can be used for model validation and calibration, and can also be applied to related problems of system stability, detection of forced oscillations, generation re-dispatch, as well as to the estimation of the system state.
\end{abstract}


% Use this to place sponsorships
\thanksto{The work was supported by funding from the U.S. DOE/OE as part of the DOE Grid Modernization Initiative.}

\vspace{-0.39cm}
\section{Introduction}
Ensuring stable, secure and reliable operations of the power grid is a primary concern for system operators \cite{kundur1994power}. Security assessment and control actions heavily rely on the accuracy of the assumed power system model and its parameters and of the estimated state \cite{sauer2017power}. Thus, inaccuracies in state estimation data or in the networked dynamic model can impact the assessment of the system stability and the efficacy of the corresponding control measures. In this paper, we explore the possibility to leverage the proliferation of Phasor Measurement Units (PMUs) that collect time synchronous data in a distributed way, for validating the assumed power system model and the current system state. In particular, our goal is to develop a data-efficient learning framework for performing an online reconstruction of the dynamic model using the minimal number of assumptions and exclusively relying on the PMU measurements. 

A number of recent works showed promising results in attacking this problem \cite{huang2009application,zhou2011calibration,guo2014adaptive,zhou2015dynamic,chen2016measurement,chavan2017identification,wang2017pmu}. Here, we propose to extend the scope of existing works to the problem of extracting the dynamic state matrix from PMU measurements in a purely data-driven way, without assuming any knowledge of model parameters. We take advantage of the separation of scales that exists in the regime of ambient fluctuations around the steady state leading to power system dynamics excited by stochastic load variations. Under quite general and widely accepted assumptions in this ambient regime, we develop a provably consistent maximum likelihood based method that recovers the dynamic state matrix with a low number of observations. Importantly, the proposed methodology can be naturally extended to cases of unknown network topology and partial observations, and has a low computational complexity which is conducive for real-time estimations.

An accurate estimation of the dynamic state matrix has a large number of applications that have been well explored in the literature \cite{machowski1997power,chiang2011direct}, including model validation and parameter calibration \cite{huang2009application,zhou2011calibration}, probing the proximity to instability and helping in design of the corresponding emergency control actions \cite{ghanavati2016identifying,van1998voltage}, optimization and resource allocation \cite{poolla2017optimal,deka2017acc}, as well as identification and analysis of forced oscillations in the system \cite{mendoza2016applying}. The potential ability to use the learned dynamic parameters to simultaneously perform a purely measurement-based state estimation of deviations in power consumption from nominal values represents another attractive feature of our framework. A validated state estimation can improve resource allocation for generation re-dispatch.

The paper is organized as follows: in Section \ref{sec:Formulation} we formulate the model and the reconstruction problem; in Section \ref{sec:Estimators} we state our learning method and discuss the convergence properties of the proposed algorithm; in Section \ref{sec:Numerics} we illustrate our approach on a test system, and provide an empirical assessment of the performance of our algorithms; finally, in Section \ref{sec:Discussion} we discuss possible extensions of our method and state some open problems.

\section{Problem formulation}
\label{sec:Formulation}

We model the power network by a graph $G=(\mathcal{V},\mathcal{E})$ with a set of $N$ nodes (buses) $\mathcal{V}$ and a set of edges (transmission lines) $\mathcal{E} \subseteq \mathcal{V} \times \mathcal{V}$. We consider the regime of ambient oscillations around the steady state that is governed by the dynamics of generator angles. It is common to model ambient dynamics with a classical equivalent model of aggregated generators \cite{chow2013power} that corresponds to a network-reduced power system where passive loads are eliminated via Kron reduction \cite{dorfler2013kron}. Although this modeling choice is not necessary for our analysis, it facilitates a uniform mathematical description where we can assume that every node $i$ in $\mathcal{V}$ essentially corresponds to a generator with non-zero inertia $M_i$ and damping $D_i$ coefficients with temporal evolution governed by the swing equations \cite{kundur1994power}:
\begin{equation}
M_i\ddot{\theta}_i+D_i(\dot{\theta}_i-\omega^{(0)})=P^{(m)}_i-P^{(e)}_i,
\label{eq:true_model}
\end{equation}
where $\omega^{(0)}$ is the synchronous frequency (60 Hz in U.S.A.); $\theta_i$ and $\dot{\theta}_i(=\partial \theta_i / \partial t)$ respectively correspond to the generator rotor angles and speeds; $P^{(m)}_i$ is the net power injection (e.g. the generator mechanical power input); and $P^{(e)}_i$ is the electrical power output. $P^{(e)}_i$ can be further expressed as a sum of power flows out of node $i$: $P^{(e)}_i = \sum_{(ij) \in \mathcal{E}}P_{ij}$, with
\begin{equation}
P_{ij} = \vert V_i\vert \vert V_j\vert \left( g_{ij} \cos (\theta_i-\theta_j) + b_{ij} \sin (\theta_i-\theta_j) \right),
\label{eq:power_flow}
\end{equation}
where conductance $g_{ij} > 0$ and susceptance $b_{ij} > 0$ correspond to the real and imaginary parts of the complex admittance $y_{ij} = g_{ij} - \hat{j} b_{ij}$ ($\hat{j}^2 = -1$) associated with each line $(i,j) \in \mathcal{E}$ in the network.

In the vicinity of the synchronous state, the difference of rotor angles is typically small, so that $\theta_i-\theta_j$ is close to zero for every pair $(i,j) \in \mathcal{E}$. Therefore, in the regime of moderate ambient fluctuations it is standard \cite{kundur1994power} to linearize the expression \eqref{eq:power_flow} around the current operating point using $\cos (\theta_i-\theta_j) \approx 1$ and $\sin (\theta_i-\theta_j) \approx (\theta_i-\theta_j)$. Over the period of time where the voltage magnitudes can be approximately considered as constant,  line admittances are characterized by effective susceptances $\beta_{ij} = \vert V_i\vert \vert V_j\vert b_{ij}$ and conductances $\gamma_{ij} = \vert V_i\vert \vert V_j\vert g_{ij}$, that absorb constant voltage magnitudes by definition. Given these simplifications due to DC linearization, we assume that the following relation is valid in expectation in the steady state regime:
\begin{equation}
\mathbb{E} P^{(m)}_i = \mathbb{E} P^{(e)}_i = \sum_{(ij) \in \mathcal{E}}\left(\gamma_{ij} + \beta_{ij} (\theta^{(0)}_i-\theta^{(0)}_j)\right),
\label{eq:steady_state}
\end{equation}
where $\theta^{(0)}_i$ denote the mean steady state values of rotor angles, and deviations of phase from these values $\delta_i = \theta_i - \theta^{(0)}_i$ are small. Note that conductances $g_{ij}$ are typically negligible for power transmission lines and hence $\gamma_{ij}$ is usually omitted in the expression \eqref{eq:steady_state} under the assumption of purely inductive lines \cite{kundur1994power}.

Finally, the resulting dynamic model that we consider in this paper takes the following form:  
\begin{equation}
M_i\dot{\omega}_i + D_i\omega_i = -\sum_{(i,j) \in \mathcal{E}}\beta_{ij}(\delta_i-\delta_j) + \delta P_i,
\label{eq:model}
\end{equation}
where $\delta P_i = P^{(m)}_i - P^{(e)}_i$ represents the effect of exogenous power deviations, and $\omega_i = \dot{\delta}_i$ denotes the relative generator rotor speed measured with respect to the reference synchronous frequency $\omega^{(0)}$.  

From \eqref{eq:model}, equations for the whole system can be written in the matrix form as
\begin{equation}
\begin{bmatrix}
\underline{\dot{\delta}}\\
\underline{\dot{\omega}}
\end{bmatrix}
=
\left[
\begin{array}{c|c}
0_{N \times N} & \mathbb{I}_{N \times N} \\
\hline
-M^{-1} L & -M^{-1} D
\end{array}
\right]
\begin{bmatrix}
\underline{\delta}\\
\underline{\omega}
\end{bmatrix}
+
\begin{bmatrix}
\underline{0}\\
M^{-1} \underline{\delta P}
\end{bmatrix},
\label{eq:model_matrix_form}
\end{equation}
where $\underline{y} = [y_1, \ldots, y_N]^{T}$ denotes a generic $N$-component vector and $y$ can refer to $\theta$, $\dot{\theta}$, $\omega$, $\dot{\omega}$ and $\delta P$; $\underline{0}$ is a $N$-dimensional vector; and $0_{N \times N}$ and $\mathbb{I}_{N \times N}$ denote $N \times N$ zero and identity matrices, respectively. $L$ is the susceptance-weighted Laplacian matrix defined as $L_{ij} = -\beta_{ij}$ for $(ij) \in \mathcal{E}$; $L_{ii} = \sum_{(ik) \in \mathcal{E}} \beta_{ik}$; and $L_{ij} = 0$ otherwise. Finally, $M$ and $D$ respectively represent diagonal inertia and damping matrices parametrized by $M_i$ $D_i$. For compactness, let us rewrite the system \eqref{eq:model_matrix_form} as
\begin{equation}
\dot{X}_t = A_{d} X_t + \tilde{\xi}_{t},
\label{eq:continuous_model}
\end{equation}
where $X_t$ is a shorthand notation for the system state vector $[\underline{\delta},\underline{\omega}]^{\top}$ at time $t$, and $\tilde{\xi}_{t}$ is the $\underline{\delta P}$ dependent vector.

\begin{figure}[!ht]
\centering
\includegraphics[width=0.97\columnwidth]{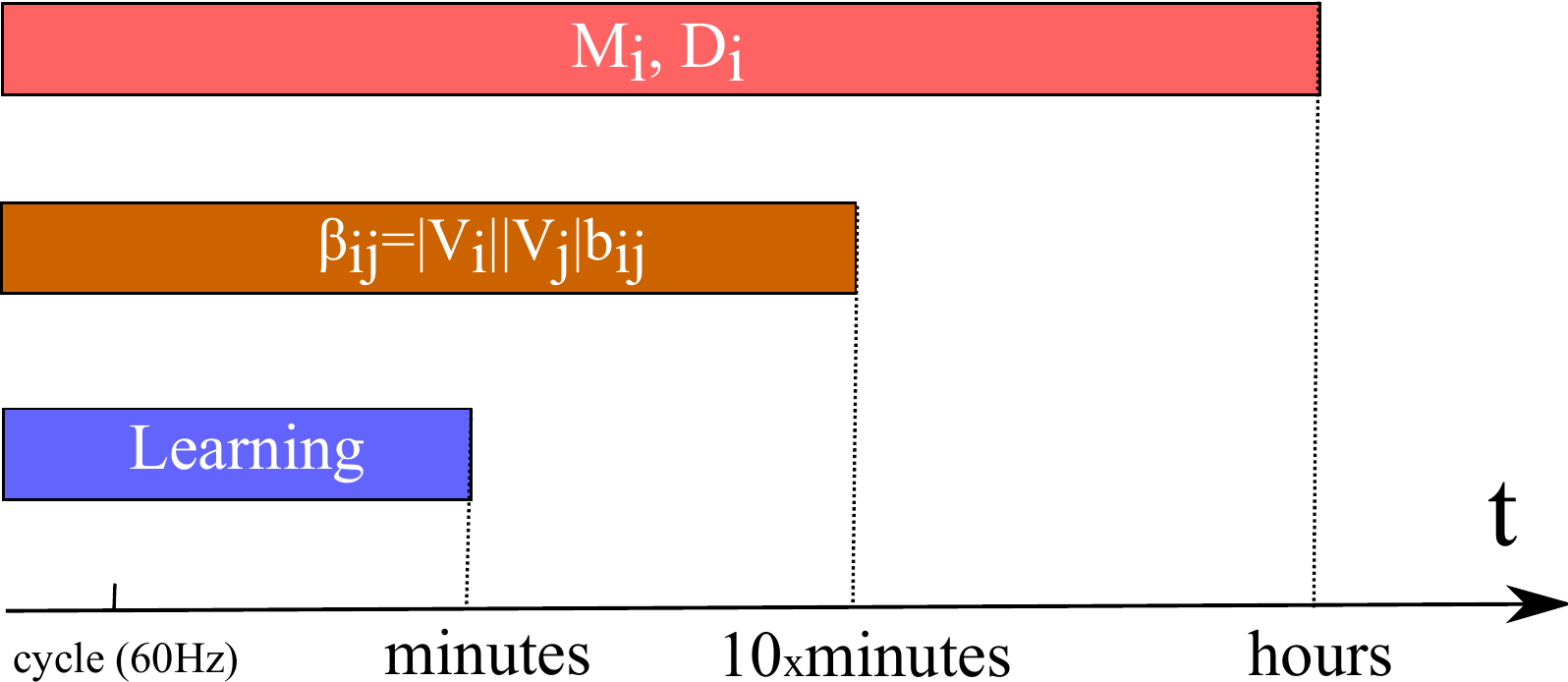}
\caption{Separation of scales in our estimation problem in the setting of ambient fluctuations. Learning procedure should be sample-efficient and only use the amount of data observed during the time period which is no longer than minutes or tens of minutes, the typical scale over which $A_{d}$ is stable.}
\label{fig:Scales}
\end{figure}

In this paper, our goal is to estimate the dynamic state matrix $A_{d}$ from PMU data providing time series measurements of dynamic variables $X_t$. In principle, $A_{d}$ can be computed if all parameters entering \eqref{eq:true_model} are known. As motivated above, here we deliberately pursue a purely data-driven approach that will provide a characterization of the system dynamic  based on the solution of the inverse problem using the observed time series, and hence would allow to validate the assumed dynamic model. However, the dynamic state matrix can be assumed to be constant only at time scales for which the parameters entering \eqref{eq:model_matrix_form} remain unchanged. In the ambient fluctuations setting, there exists a natural separation of time scales in parameter variations, see Figure~\ref{fig:Scales}. The parameters of generators, such as inertia $M_i$ or damping coefficients $D_i$ can be regarded as stable on the scale of several hours, with a potential slow drift due to droop or local feedback control \cite{sauer2017power}. Although infrequent changes in the system may provoke a change in $\vert V_i\vert$, overall voltage magnitudes may be considered stable on the time interval of the order of tens of minutes \cite{kundur1994power}. The same conclusion holds for the values $b_{ij}$ that may fluctuate due to variations in temperatures that rarely happen on shorter scales compared to tens of minutes. On the other hand, we assume that ambient fluctuations themselves are caused by fluctuations of loads $P^{m}_i$ and $P^{e}_i$ around base nominal values or generator noise \cite{wang2017pmu}, and therefore occur at very short scales of the order of the frequency cycle of $60$ Hz. Because of this reason and due to the aggregated nature of loads, the total power deviations $\delta P_i$ are commonly modeled as random Gaussian variations \cite{singh2010statistical}. Therefore, learning must take place using measurements obtained during the interval of time that happens on the scale of minutes, below the scale on which voltage magnitudes might change, and above the scale of load fluctuations $\delta P_i$. We model $\delta P_i$ as a zero mean Gaussian noise term with standard deviation $\sigma_{P_i}$ that incorporates the aggregated ambient fluctuations in power injection and consumption.

PMU measurements are discrete in nature, and arrive as time-separated data samples with a typical frequency of several cycles. Therefore, the observed data points approximately follow the dynamics representing the discretization of (\ref{eq:continuous_model}) with a certain step $\Delta t$. Using the Euler-Maruyama discretization scheme, we get to the first order in $\Delta t$:
\begin{equation}
X_{t+1} = A X_{t} + B\xi_{t},
\label{eq:discrete_model}
\end{equation}
with $A = (A_{d} \Delta t + \mathbb{I}_{2N \times 2N})$, $\xi_{t} \sim \mathcal{N}(\underline{0},I_{2N \times 2N})$ is the standard multivariate normally distributed noise and $B$ summarizes the resulting scale of fluctuations. It is reasonable to assume that load fluctuations are spatially independent across nodes so that $B$ is diagonal; however, variance at individual buses can be different, so that $B_{ii} = M^{-1}_i \sigma_{P_i} \sqrt{\Delta t}$ for $i \in [N+1,2N]$ has a meaning of the noise standard deviation at node $i$ and $B_{ii} = 0$ for $i \in [1,N]$. As we will see below, $\Delta t$ is an important parameter that drives the reconstruction procedure. Indeed, $\Delta t$ can not be smaller than the resolution of the PMU data, and should not be too small so that $\xi_{t}$ could be conveniently interpreted as uncorrelated white noise across time. At the same time, $\Delta t$ can also not be too large because in this case $X_{t}$ would be essentially independent across time, meaning that the dynamic state matrix $A_{d}$ could not be recovered in principle, and one could only hope to get the estimation of the steady state covariance \cite{deka2017state}. In order to facilitate the learning task, it is advantageous to select $\Delta t$ in such a way that it satisfies the trade-off between the amount of observations used and the accuracy achieved. In what follows and unless stated otherwise, we will assume that $\xi_{t}$ is independent and identically distributed for the purpose of reconstruction of the dynamic state matrix $A_{d}$.

% \begin{figure}[!ht]
% \centering
% \includegraphics[width=\columnwidth]{pscc_abstract_image1.pdf}
% \caption{Schematic representation of the problem: given phases and frequencies of PMU measurements at a subset of buses in the power grid under ambient fluctuations, we learn the parameters of lines and generators.}
% \label{fig:Formulation}
% \end{figure}

\section{Estimators}
\label{sec:Estimators}

\subsection{Maximum likelihood formulation}

In this section, we present our estimators for the dynamic state matrix from $T$ discrete observations of the system $\{X_{t}\}_{t=1,\ldots,T}$. Consider the empirical cross-correlation matrices with and without displacement that respectively read
\vspace{-0.1cm}
\begin{align}
    \Sigma_{1} &= \frac{1}{T-1}\sum\limits_{t=1}^{T-1} X_{t+1}X^{\top}_{t}, \label{eq:cross-corr1} \\ 
    \Sigma_{0}   &= \frac{1}{T-1}\sum\limits_{t=1}^{T-1} X_{t}X^{\top}_{t}. \label{eq:cross-corr0}
\end{align}
Based on Eq.~\eqref{eq:discrete_model} we introduce the following estimator for the matrix $A$ which exists whenever the cross-correlation matrix in Eq.~\eqref{eq:cross-corr0} is invertible:
\begin{equation}
\widehat{A} = \Sigma_{1} \Sigma_{0}^{-1}. \label{eq:S_estimator}
\end{equation}
We refer to this estimator as to the Maximum Likelihood (ML) estimator for the $A$. Notice that the $2N\times 2N$ matrices $\Sigma_{0}$ and $\Sigma_{1}$ are at most of rank $T-1$ as expressions \eqref{eq:cross-corr1} and \eqref{eq:cross-corr1} are sums of $T-1$ rank one matrices. This implies that the matrix $\Sigma_{0}$ is not invertible for $T\leq 2N$. When $T\geq 2N+2$, the matrix is invertible with probability one since $\text{rank}\left\{X^{\top}_{T-1},\ldots,X^{\top}_1\right\} = \text{rank}\left\{\xi^{\top}_{T-2},\ldots,\xi^{\top}_{1},X^{\top}_1\right\} \geq 2N$ and $\xi_{t}$ are independent normally distributed vectors.

Bounds on the expected reconstruction error for the ML estimator are given by the following theorem.

\begin{thm}[Reconstruction error for discrete dynamics]\label{th:s-error}
Let $\epsilon>0$ and $T>2N+2$. The ML estimator~\eqref{eq:S_estimator} reconstructs the matrix $A$ with probability at least $1-\epsilon$ within a Frobenius norm error bounded by
\begin{align}\label{eq:shity_bound}
   \Vert \widehat{A} - A \Vert_{\text{F}} \leq \frac{\Vert B \Vert_{2}}{\epsilon \sqrt{T-1}}  \sqrt{\mathbb{E}\left[\Tr(\Sigma_0)\right]\mathbb{E}\left[\Vert\Sigma^{-1}_0\Vert_{\text{F}}^2 \right]}.
\end{align}
\end{thm}
\begin{IEEEproof} 
The proof is given in Appendix~\ref{app:theorem}.
\end{IEEEproof}

The bound in Theorem~\ref{th:s-error} is valid without any assumptions on the matrix $A$. For a stable system with a steady state dynamics one expects that $\Sigma_0$ concentrates to its expectation as $T\rightarrow \infty$. In this case Theorem~\ref{th:s-error} shows that the error on the ML estimator decreases as the inverse square-root of the number of observations and increases linearly with the noise intensity.
%For more refined error-bounds in the steady-state dynamics we refer the reader to \cite{lutkepohl1993introduction}.

The dynamic state matrix describing the continuous dynamics in Eq.~\eqref{eq:continuous_model} is estimated from the discrete dynamic matrix using the relation
\begin{equation}
\widehat{A}_d = \frac{\widehat{A} - \mathbb{I}_{2N\times2N}}{\Delta t}. \label{eq:discrete_to_continuous}
\end{equation}
Guarantees on the reconstruction error for the continuous dynamic matrix follow from Theorem~\ref{th:s-error}.
\begin{cor}\label{co:continuous_dyn}
Let $\epsilon>0$ and $T>2N+2$. The error on the reconstruction of $A_d$ is with probability at least $1-\epsilon$ upper-bounded by
\vspace{-0.2cm}
\begin{align}\label{eq:dyn_bound}
   \Vert \widehat{A}_{d} - A_{d} \Vert_{\text{F}} \leq \epsilon^{-1}\sqrt{\frac{\sum_{i=1}^{N}M_{i}^{-2}\sigma_{P_i}^2}{\Delta t(T-1)}\mathbb{E}\left[\Tr(\Sigma_0)\right]\mathbb{E}\left[\Vert\Sigma^{-1}_0\Vert_{\text{F}}^2 \right]}.
\end{align}
\end{cor}
\begin{IEEEproof} 
It is a direct application of Theorem~\ref{th:s-error} with $B$ satisfying $B_{ii} = M^{-1}_i \sigma_{P_i} \sqrt{\Delta t}$ for $i \in [N+1,2N]$.
\end{IEEEproof}

The main implication of Corollary~\ref{co:continuous_dyn} is that the error on $A_d$ decreases with respect to the product $T \Delta t$. This product corresponds to the total observation time of the system $t_{\text{obs}}$. Therefore, if the number of data samples $T$ is large enough for $\Sigma_0$ to concentrate to its average, the reconstruction error only depends on $t_{\text{obs}} = T \Delta t$ through its inverse square root.
%was: the reconstruction error does not depend on the total observation time.
Notice that for a fixed observation time, the error does not depend on the discretization, see \cite{bento2010learning} for an extended discussion on this property. 

As mentioned earlier, \eqref{eq:S_estimator} can be interpreted as the maximum likelihood estimator for the dynamic matrix $A$. This means that the estimator $\widehat{A}$ can be seen as the outcome of some minimization procedure. While obtaining $\widehat{A}$ through an optimization problem might seem unnecessarily more complicated, formulating the estimator as a minimization procedure renders possible the incorporation of extra information in the estimation. This extra information can take the form of a prior on the inertia, damping parameters or line susceptances in the system, or it can serve as a prior on the location of zero elements in $A$. As it is crucial to keep the learning time below the typical time for which system parameters drift (see Figure~\ref{fig:Scales}), adding extra information in the reconstruction help in accomplishing this task by increasing the learning accuracy for a fixed observation period. The precise minimization procedure is given by the following proposition.

\begin{prop}[Maximum-Likelihood estimator]\label{pr:pipeau}
Given $T$ observations $\{X_{t}\}_{t=1,\ldots,T}$ resulting from the discrete linear dynamics \eqref{eq:discrete_model}, the maximum likelihood estimator of $A$ represents the solution of the following least-squares regression
\vspace{-0.1cm}
\begin{equation}
\widehat{A} = \argmin_{A} \sum\limits_{t=1}^{T-1} \Vert X_{t+1} - A X_{t} \Vert_{2}^{2}.
\label{eq:objective}
\end{equation}
Moreover, for $T\geq 2N+2$, the minimum of the least-square regression is achieved by $\widehat{A}=\Sigma_1\Sigma_0^{-1}$.
\end{prop}

\begin{IEEEproof} 
The proof is given in Appendix~\ref{app:prop}.
\end{IEEEproof}

Notice that the first $N$ rows of the dynamic state matrix $A_{d}$ in Eq.~\eqref{eq:model_matrix_form} are $\left[ 0_{N\times N} \quad \mathbb{I}_{N \times N} \right]$. Even though it seems natural to incorporate this information in the estimation procedure, in fact it appears to be unnecessary as these rows are not directly affected by noise (given that $B_{ii} = 0$ for $i \in [1,N]$) and are always reconstructed perfectly. However the situation is different for the diagonal lower-right block of $A$, $\left[ M^{-1} D \right]$ that is subject to noise. Thanks to the optimization formulation in Eq.~\eqref{eq:objective}, we can restrict the regression to matrices $A$ that have a diagonal lower-right $N\times N$ block to obtain the following constrained estimator,
\vspace{-0.1cm}
\begin{align}
\widehat{A} = &\argmin_{A} \sum\limits_{t=1}^{T-1} \Vert X_{t+1} - A X_{t} \Vert^{2}_{2} \nonumber\\
&\text{s.t.} \enskip A_{ij} = 0 \enskip \forall i,j \in [N+1,2N] \enskip \text{and} \enskip i\neq j. \label{eq:CML}
\end{align}
While the constrained estimator \eqref{eq:CML} provides a more accurate reconstruction than its unconstrained counterpart \eqref{eq:S_estimator}, it is computationally more expensive as it requires to solve an optimization problem. This trade-off between computational power and accuracy can be crucial for practical applications. 

\subsection{Extensions}

Other extensions of the ML estimator can be considered based on prior information that we can incorporate in the least-squares regression \eqref{eq:objective}.
For instance, when the system parameters are expected to drift slowly it is beneficial to incorporate a prior on the matrix $A$. This prior can take the form of a Gaussian prior centered around a previous reconstruction. This translates into a least-squared regression with a Tikhonov regularization \cite{Vauhkonen1998Tikhonov}, i.e.
\begin{equation}
\widehat{A} = \argmin_{A} \left( \sum\limits_{t=1}^{T-1} \Vert X_{t+1} - A X_{t} \Vert^{2}_{2} + \nu \Vert A - \widehat{A}_{\text{prev}} \Vert_{2}^{2} \right),
\end{equation}
where $\nu$ is a parameter indicating our degree of confidence that the current matrix is close to its previous estimate $\widehat{A}_{\text{prev}}$. 

There exists instances when the system parameters might not have been previously estimated, for example when the topology has been modified due to outages, lines tripping or controller changes \cite{wang2017pmu}, but the topology of the observed grid is known to be sufficiently sparse. In this case it would be beneficial to promote the sparsity of the matrix $A$ with a Laplace prior $P(A) \propto e^{-\lambda \Vert A \Vert_{1}}$. This prior leads to the following LASSO estimator
\vspace{-0.1cm}
\begin{equation}
\widehat{A} = \argmin_{A} \left( \sum\limits_{t=1}^{T-1} \Vert X_{t+1} - A X_{t} \Vert^{2}_{2} + \lambda \Vert A \Vert_{1} \right).
\end{equation}
The LASSO estimator proves to be much more efficient than the unconstrained least-squares \eqref{eq:objective} when the matrix is sparse, rendering the sample requirement very weakly dependent of the size of the problem $N$ \cite{bento2010learning}.

Finally in cases where the state of only a subset of buses is observed \cite{deka2017state}, it is still possible to retrieve part $A$ of the dynamic state matrix corresponding to the visible part of the system. This can be done with the so-called ``sparse plus low-rank" heuristic producing $\widehat{A}$ and $\widehat{L}$, the minimizers of

\vspace{-0.39cm}
\begin{equation}
\sum\limits_{t=1}^{T-1} \Vert X_{t+1} - (A+L) X_{t} \Vert^{2}_{2} + \lambda \Vert A \Vert_{1} + \eta \Vert L \Vert_{*},
\end{equation}
where $L$ summarizes the effects of the unobserved measurements and the nuclear norm penalty represents a convex surrogate for the low rank constraint, see \cite{jalali2011learning} for more details.

% \begin{align}
% &(\widehat{A}, \widehat{L})= \argmin_{A,L}  \nonumber
% \vspace{-0.2cm}
% \\& \left( \sum\limits_{t=1}^{T-1} \Vert X_{t+1} - (A+L) X_{t} \Vert^{2}_{2} + \lambda \Vert A \Vert_{1} + \eta \Vert L \Vert_{*} \right),
% \end{align}

%Under some identifiability conditions, in the limit $M \to \infty$ the minimizer will be given by $\widehat{A} = A_{\mathcal{OO}}$ and $\widehat{L} = A_{\mathcal{OH}} \Sigma_{\mathcal{HO}} \Sigma_{\mathcal{HH}}^{-1}$

%Dynamics converges to a Gaussian distribution with covariance
%\begin{equation*}
%\Sigma =
%\begin{bmatrix}
%    \Sigma_{\mathcal{OO}} & \Sigma_{\mathcal{OH}} \\
%    \Sigma_{\mathcal{HO}} & \Sigma_{\mathcal{HH}}
%\end{bmatrix}
%\end{equation*}

\section{Case study}
\label{sec:Numerics}

\begin{figure}[!ht]
\centering
\includegraphics[width=0.97\columnwidth]{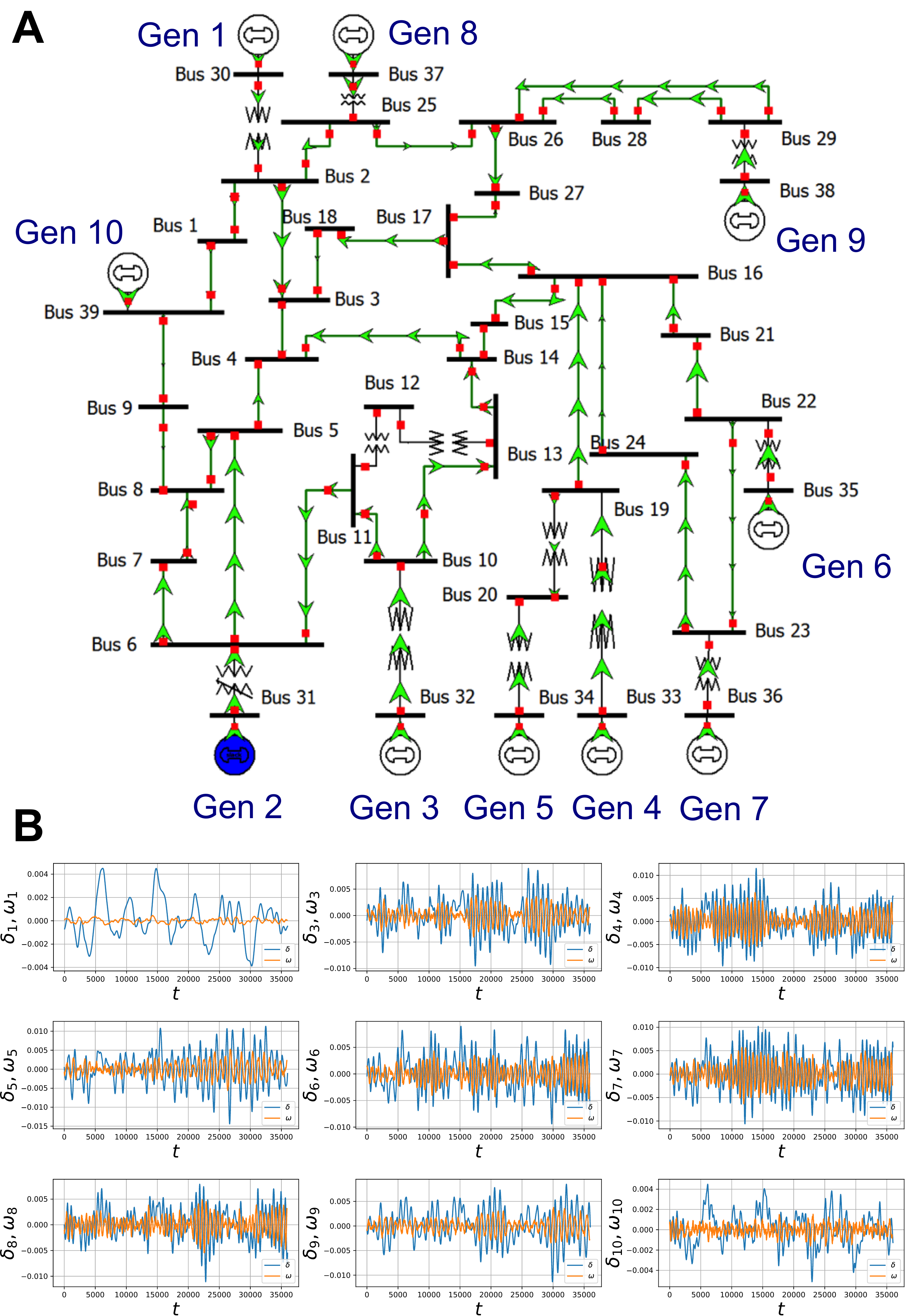}
\caption{(A) Topology of the IEEE 39-bus system with 10 generators \cite{demetriou2015dynamic}. (B) Synthetic PMU-measured data generated using the parameters of the Kron-reduced system and sampled at $60$ Hz frequency. 10 minutes of data is presented for every generator except generator 2 that serves as the reference slack bus. The ``time'' $t$ counting the number of sample points is expressed in terms of the number of cycles. Note that the time series for the generator 1 has a distinct appearance due to a significantly larger interia coefficient $M_{1}$.}
\vspace{-0.5cm}
\label{fig:IEEE39}
\end{figure}

We illustrate the performance of our learning algorithms on the IEEE 39-bus 10-generator test system \cite{demetriou2015dynamic}; the topology of this system is shown in Figure~\ref{fig:IEEE39}~(A). We assume that PMUs are located at the generator buses. First, we perform Kron reduction and eliminate passive loads, obtaining the parameters of the system \eqref{eq:model_matrix_form} obeyed by the generators \cite{wang2017pmu}. Given the established parameters of the reduced network and hence the ground truth $A_{d}$, we use the discretized representation \eqref{eq:discrete_model} to simulate the dynamics and produce the time series on $\delta_i$ and $\omega_i$ for each generator $i \in \mathcal{V}$ using the smallest resolution $\Delta t = 1/60$ sec (1 cycle) for which our model \eqref{eq:model} is still valid; in Figure~\ref{fig:IEEE39}~(B), we show one example of such run with the simulated data over $10$ minutes. In all simulations, the load variation is fixed at the level of $\sigma_{P_i} = 0.01$ p.u. \cite{wang2017pmu}. We use data obtained in this way in all reconstruction experiments reported below.

Notice that PMUs might have a lower time resolution, subsampling these data points with a different time step $\Delta t$, for instance once every $k$ cycles (for many measuring devices, the maximum sampling rate corresponds to $k=2$ or $k=3$). Moreover, because of the data processing reasons, one might wish to deliberately sample data at a lower frequency for the reconstruction purposes. Therefore, it is instructive to check the sensitivity of the algorithmic performance to the chosen subsampling step $\Delta t$. But prior to that, we need to establish the measure of performance for the two estimators introduced in this paper: the Unconstrained Maximum Likelihood (UML) estimator \eqref{eq:S_estimator} and the Constrained Maximum Likelihood (CML) estimator \eqref{eq:CML}, where the word ``constrained'' means that the support pattern of $A_{d}$ has been explicitly enforced. We use both algorithms to produce an estimate $\widehat{A}_{d}$ of the dynamic state matrix $A_{d}$. For a given time separation $\Delta t$ between measurement samples $\{X_t\}$, first the discrete matrix $\widehat{A}$ at the corresponding scale \eqref{eq:discrete_model} is estimated, and then the dynamic state matrix is recovered using the linear approximation \eqref{eq:discrete_to_continuous}.

In order to account for the additional sparsity structure present in $A_{d}$, we supplement the application of \eqref{eq:S_estimator} in the UML estimator with a post-estimation thresholding of matrix elements that are known to be zero, in particular in the block corresponding to the diagonal submatrix $M^{-1}D$ in \eqref{eq:model_matrix_form}. We quantify the quality of the estimation using the relative Frobenius error $\varepsilon$, defined as
\begin{equation}
    \varepsilon = \frac{\Vert \widehat{A}_{d} - A_{d} \Vert_{F}}{\Vert A_{d} \Vert_{F}},
\end{equation}
where $\widehat{A}_{d}$ is the recovered matrix and $A_{d}$ is the ground truth dynamic state matrix used to produce data (at finer discretization $\Delta t = 1/60$ sec).

\begin{figure}[!ht]
\centering
\includegraphics[width=0.97\columnwidth]{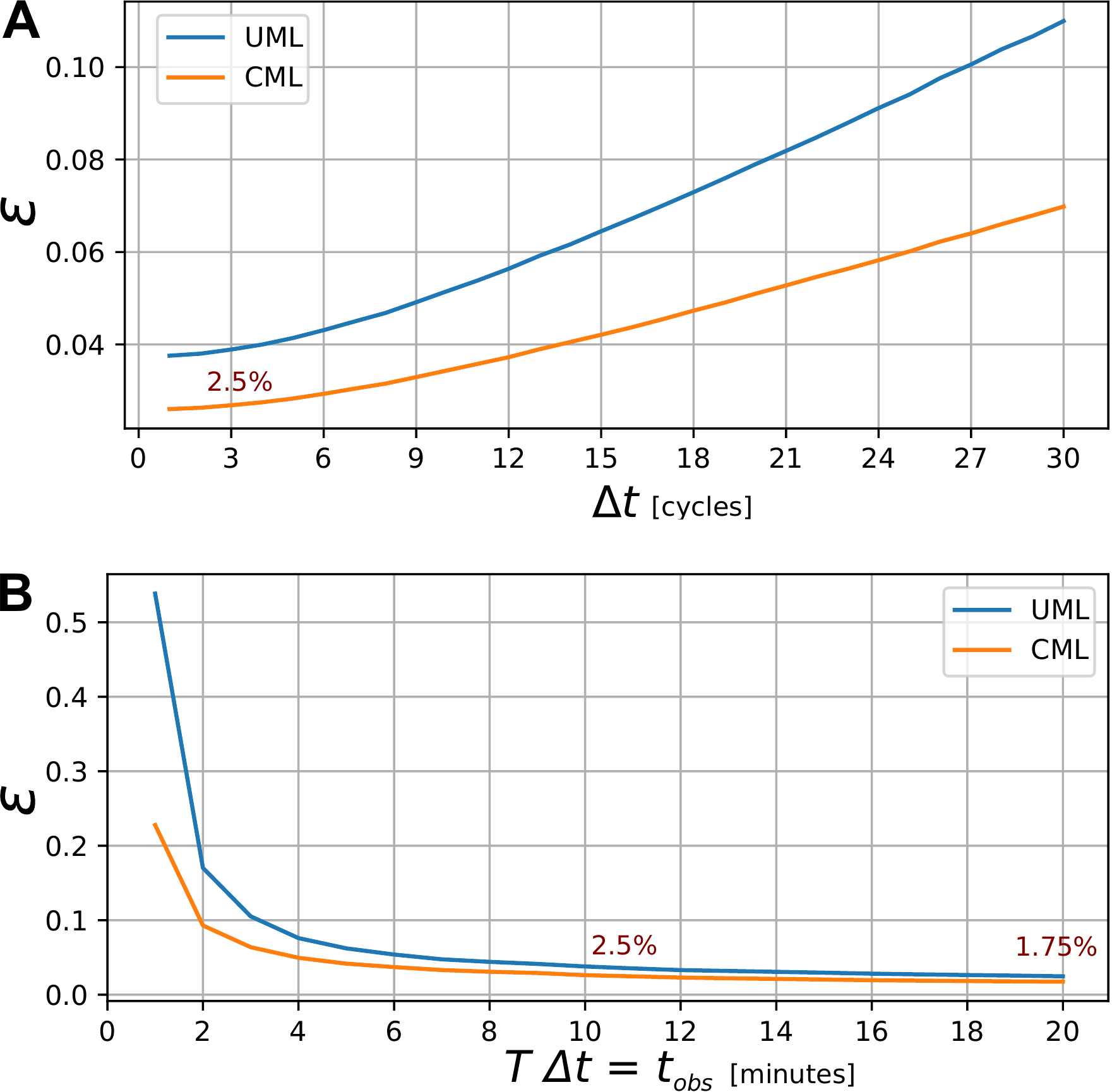}
\caption{Performance of the estimators as a function of (A) time discretization $\Delta t$ and (B) total observation time $t_{\text{obs}} = T \Delta t$. Each data point is averaged over $50$ independent realizations of the dynamics for reducing the noise due to statistics.}
\label{fig:Fig3}
\end{figure}

The dependence on $\Delta t$ is shown in Figure~\ref{fig:Fig3}~(A). In this figure, the observation window $t_{\text{obs}}$ is fixed to $10$ min, and therefore the number of samples $T$ seen by the algorithms decreases with $\Delta t$ as $T = t_{\text{obs}} / \Delta t$. According to the Corollary 2, the reconstruction error $\varepsilon$ should stay constant in $T \Delta t = t_{\text{obs}}$. We see that for both algorithms we don't see any clear plateau in $\varepsilon(\Delta t)$ dependence even for small $\Delta t$. This is due to the increasing with $\Delta t$ error in the first-order approximation \eqref{eq:discrete_model} of the fine-grained dynamics and hence to the level of validity of \eqref{eq:discrete_to_continuous}. Nevertheless, we see that for both estimators $\varepsilon$ is growing slowly with $\Delta t$, and the error seen for $\Delta t = 3$ cycles is very close to the one realized when the finest possible discretization $\Delta t = 1$ is taken, although the algorithms use three times less samples in the former case. Moreover, given that $\Delta t = 3$ cycles represents a normal sampling rate for many measurement devices, we use this sampling rate in the subsequent experiments. The final thing that we observe is that the algorithm CML that exploits the structure of $A_{d}$ systematically yields a lower error compared to UML even with the added heuristic post-reconstruction thresholding; on the other hand, it should be noted that CML is slower as one needs to solve the optimization problem \eqref{eq:CML} instead of just inverting the matrix (in the present experiments, the optimization was carried out using the Ipopt solver). Still, both algorithms run in seconds for this test case, which represents a premise for an online real-time implementation.

In Figure~\ref{fig:Fig3}~(B), we study the performance of our estimator as a function of the number of samples $T$ for a fixed sampling step $\Delta t = 3$ cycles (and hence for growing observation time $t_{\text{obs}}$ from $1$ to $20$ minutes). We see that the experiment confirms the conclusion of the Corollary 2: the estimated dynamic state matrix $\widehat{A}_{d}$ quickly converges to the ground truth matrix $A_{d}$, with CML algorithm achieving the relative error of $2.5 \%$ by $t_{\text{obs}} = 10$ min and $1.75 \%$ by $t_{\text{obs}} = 20$ min. This fact shows that it is possible to estimate the dynamic state matrix to an impressive accuracy under the time constraints outlined in Figure~\ref{fig:Scales}.

\begin{figure}[!ht]
\centering
\includegraphics[width=0.97\columnwidth]{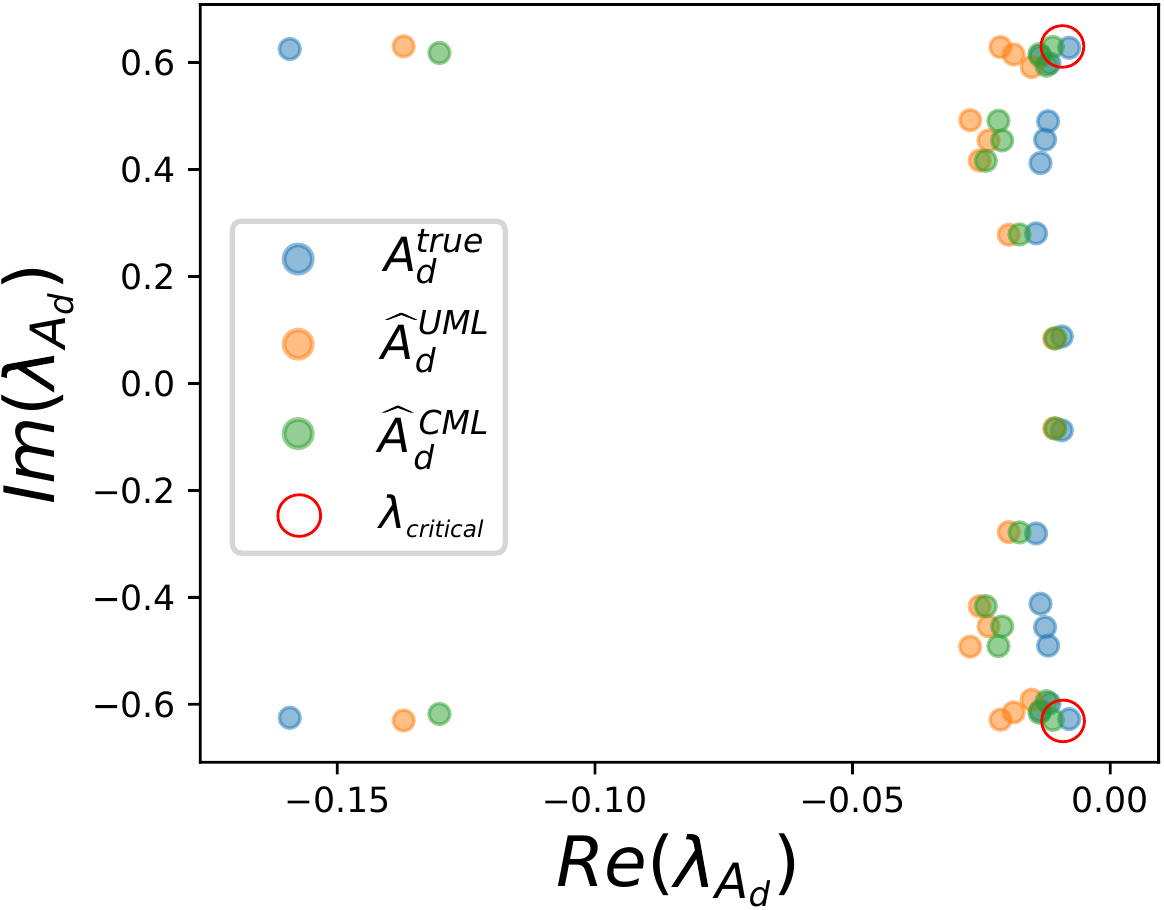}
\caption{Quality of prediction of the critical eigenvalues of the dynamic state matrix $A_d$ with $t_{\text{obs}}=10$ min and $\Delta t = 3$ cycles.}
\label{fig:Fig4}
\end{figure}

Besides the accurate prediction of the dynamic state matrix $A_{d}$ per se, a desirable feature of the estimators would consist in an accurate prediction of the properties of this matrix, in particular including the spectral properties \cite{wang2017pmu}. Indeed, the critical eigenvalue  is known to serve as measures of proximity to the instability \cite{machowski1997power,ghanavati2016identifying,van1998voltage}, while the associated critical eigenvector might provide useful information on the system response and facilitate the design of control actions such as the generation re-dispatch \cite{ghanavati2016identifying,mendoza2016applying}. In Figure~\ref{fig:Fig4}, we test the accuracy of the critical eigenvalues prediction using the samples obtained within the $t_{\text{obs}} = 10$ min observation interval at the sampling rate $\Delta t = 3$ cycles. Is is apparent that the critical eigenvalues are predicted to a good accuracy, which shows that our learning procedure can be used for the online monitoring of the system stability.

%a number of realistic test cases using synthetic and real-world data. Ability to learn the model on the fly would be a highly desirable feature of the algorithm. In this regards, we show that it is possible to construct an online version of our algorithm by including a prior on the slowly evolving parameters, which leads to an additional regularization term in the optimization formulation. We also discuss an attractive feature of our framework which allows, under some mild additional assumptions, to make state estimation of nodal injections in parallel with learning the dynamic parameters.

\section{Discussion and conclusions}
\label{sec:Discussion}

In this work, we explored the maximum likelihood based approach to the problem of estimation of the dynamic state matrix from PMU measurements. In particular, we constructed and tested two least-squares estimators, one based on a fast inversion of the empirical covariance matrix, and another one based on the solution of a convex quadratic regression taking full advantage of the problem structure and leading to a more accurate reconstruction, but at an expense of a slightly higher computational time. These two estimators realize a common trade-off in practical applications between the accuracy and the speed of computations. In this contribution, we have verified the properties of our algorithms on synthetic data from standard IEEE test case; in future studies, it would be instructive to test their performance on data collected from real-world PMUs.

As clarified in section~\ref{sec:Estimators}, our framework is very broad and can naturally accommodate the regularized online learning by incorporating the previously learned models as a prior thus potentially decreasing the computational time even further, as well as extensions to the case of sparse network topologies and to realistic scenario of incomplete observations due to a partial PMU coverage \cite{deka2017state}. It would be useful to perform a theoretical finite-sample analysis of these extensions in the future similarly to the analysis presented in this paper, as well as to provide an empirical assessment of the algorithm performance in realistic problems.

As we commented while motivating the dynamic learning problem, estimated parameters can be useful in a number of tasks related to optimization, control and security of the transmission grid. Another attractive feature of our methodology consists in an ability to perform data-driven state estimation of power fluctuations through estimation of the matrix $B$ related to power fluctuations \eqref{eq:discrete_model}, as explained in the proof of the Proposition~\ref{pr:pipeau} in Appendix~\ref{app:prop}. It is important to quantify the potential advantage of using the learning-based method in these applications.

Some of  relevant open questions that we did not address in this work include the construction of optimal estimators in the cases of more general noise distributions than the Gaussian case (for example, the power-law distributed noise), as well as for noise correlated in space and time. We anticipate that in the regime of weak spatial and temporal coupling the estimators introduced in this paper should be sufficiently robust and perform reasonably well, but this statement should be thoroughly checked on realistic test cases.

Finally, an interesting and natural direction for future exploration consists in extending our scheme to non-stationarity and to higher-order dynamic models.

\appendices

\renewcommand{\theequation}{\thesection.\arabic{equation}}

\section{Proof of Theorem \ref{th:s-error}}\label{app:theorem}

First multiply both sides of Eq.~\eqref{eq:discrete_model} by $X^{\top}_{t}$ and perform a sum over $t$ to obtain
\vspace{-0.2cm}
\begin{align}\label{eq:app_bt1}
    \Sigma_1 = A \Sigma_0 + R,
\end{align}
where
\vspace{-0.39cm}
\begin{align}
    R_{ij} = \frac{1}{T-1} \sum\limits_{t=1}^{T-1} B_{ii} \xi_{t,i}X_{t,j}.
\end{align}
Since $T\geq 2N+2$, the cross-correlation matrix $\Sigma_0$ is invertible. Multiplying both sides of Eq.~\eqref{eq:app_bt1} by $\Sigma^{-1}_0$ gives $\widehat{A}-A = R\Sigma_0^{-1}$. In expectation the difference in Frobenius norm between the ML estimator and the matrix A is upper-bounded by the following expression
\begin{align}
   \mathbb{E}\left[ \Vert \widehat{A} - A \Vert_{\text{F}} \right] &= \mathbb{E}\left[\Vert R \Sigma_0^{-1} \Vert_{\text{F}} \right],\\
   &\leq \mathbb{E}\left[\Vert R \Vert_{\text{F}} \Vert \Sigma_0^{-1} \Vert_{\text{F}}\right],\\
   &\leq \sqrt{\mathbb{E}\left[\Vert R \Vert_{\text{F}}^2\right] \mathbb{E}\left[\Vert  \Sigma_0^{-1} \label{eq:add_bt2} \Vert_{\text{F}}^2\right]},
\end{align}
where in two last lines we have used Cauchy-Schwarz inequality for the Frobenius norm and for the expectation, respectively.
In order to compute the expected Frobenius norm of $R$, we first evaluate the expectations of its elements squared
\begin{align}
    \mathbb{E}\left[R_{ij}^{2}\right] &= \left(\frac{B_{ii}}{T-1}\right)^2 \sum\limits_{t_1=1}^{T-1}\sum\limits_{t_2=1}^{T-1} \mathbb{E}\left[\xi_{t_1,i}X_{t_1,j}\xi_{t_2,i}X_{t_2,j}\right],\\
    &= \left(\frac{B_{ii}}{T-1}\right)^2 \sum\limits_{t=1}^{T-1} \mathbb{E}\left[X_{t,j}^2\right], 
\end{align}
where we used that $\mathbb{E}\left[\xi_{t,i}^2\right]=1$ and $\mathbb{E}\left[\xi_{t_1,i}X_{t_1,j}\xi_{t_2,i}X_{t_2,j}\right]=0$ if $t_1 \neq t_2$. It is now easy to compute the expected Frobenius norm of the matrix $R$,
\begin{align}
    \mathbb{E}\left[\Vert R\Vert_{\text{F}}^2 \right] &= \frac{\Vert B \Vert_{2}^{2}}{T-1}  \mathbb{E}\left[\Tr \left( \Sigma_0 \right)\right].\label{eq:add_bt3}
\end{align}
The final step of the proof consists in combining Eq.~\eqref{eq:add_bt2}, Eq.~\eqref{eq:add_bt3} and Markov inequality.

\section{Proof of Proposition \ref{pr:pipeau}}\label{app:prop}

The likelihood function is the probability density function (PDF) of the observation given the parameters $A,B$ that define the model. Since $\xi_t = B^{-1}(X_{t+1} - A X_{t})$ and the noise is identically and independently distributed, we can related the PDF of ${\bf X} = X_1,\ldots,X_T$ for a given $A,B$ to the PDF of a single $\xi$, i.e.
\vspace{-0.2cm}
\begin{align}\label{eq:likelihood_pipeau}
    \rho_{X_t}({\bf X}) =\frac{\rho_{X_1}(X_1)}{\text{det}(B)} \prod_{t=1}^{T-1}\rho_{\xi}(B^{-1}(X_{t+1} - AX_{t})).
\end{align}

The maximum likelihood estimator is given by the argmax of the likelihood in Eq.~\eqref{eq:likelihood_pipeau} with respect to $A,B$. Equivalently one can minimize the opposite of the logarithm of the likelihood to obtain
\begin{align}
    (\widehat{A},\widehat{B}) = \argmin_{A,B} \left[ \ln \det B-\sum_{t=1}^{T-1} \ln \rho_{\xi}(B^{-1}(X_{t+1} - AX_{t})) \right]
\vspace{-0.2cm}
\end{align}
%Using Bayes' theorem,
%\begin{align*}
%&\mathbb{P}(\underline{X} \mid A,B) = \int d\underline{\xi} \, \mathbb{P}(\underline{X} \mid A,B,\underline{\xi})\mathbb{P}(\underline{\xi})\\
%& = \int \prod_t d\xi_t \delta(X_{t+1} - A X_{t} - B \xi_{t})\frac{1}{(2\pi)^{n/2}}\exp\left(-\frac{1}{2}\Vert \xi_{t} \Vert^{2}_{2}\right)\\
%& = \int \prod_{t,k} \frac{dz_k^{t}}{(2\pi)^{n/2}b_{k}} \delta(X_{t+1} - A X_{t} - z_{t})\exp\left(-\frac{1}{2} z^{T}_{t}B^{-2}z_{t}\right) \\
%& = \prod_{t,k} \frac{1}{b_{k}} \exp\left(- \frac{1}{2b^{2}_{k}}(X_{t+1} - A X_{t})^{2}_{k} \right)\\
%& = \prod_{k} \frac{1}{(b_{k})^{T}} \exp \left(- \frac{1}{2b_k^2} \sum_{t=1}^{T} (X_{t+1}-AX_{t})^{2}_{k} \right)
%\end{align*}
After replacing the normally distributed PDF for $\xi$, we arrive at the following optimization problem
\begin{equation}\label{eq:max_likelihood_pipeau}
(\widehat{A},\widehat{B}) = \argmin_{A,B} \sum_{i=1}^{N}\left[\ln B_{ii} + \frac{(2B^{2}_{ii})^{-1}}{T-1}\sum\limits_{t=1}^{T-1} (X_{t+1} - A X_{t})_{i}^{2} \right]
\end{equation}
Notice that in Eq.~\eqref{eq:max_likelihood_pipeau}, the optimization over $A$ is independent of the value of $B$ and can be performed separately, yielding the optimization problem \eqref{eq:objective} that after some algebra can be equivalently represented as
\vspace{-0.1cm}
% \begin{align}\label{eq:max_likelihood_pipeau_the_return}
% \widehat{A} &= \argmin_{A} \frac{1}{T-1}\sum\limits_{t=1}^{T-1} \Vert X_{t+1} - A X_{t} \Vert_{2}^{2}. %,\\
% %&=\argmin_{A} \frac{1}{T-1}\sum\limits_{t=1}^{T-1} \Tr((X_{t+1} - A X_{t})(X_{t+1} - A X_{t})^{\top}),
% \end{align}
%where in the last line we use the cyclic property of the trace. With a little algebra, the Maximum-Likelihood estimator for the dynamic state matrix can be seen as the result of the following quadratic optimization
\begin{align}\label{eq:pipeau_final_stage}
\widehat{A} &= \argmin_{A}  \Tr(A^{\top}A\Sigma_0 - 2A^{\top}\Sigma_1).
\end{align}
%where $\Sigma_1$ and $\Sigma_0$ are defined by Eq.~\eqref{eq:cross-corr1} and Eq.~\eqref{eq:cross-corr0} respectively.
Whenever $\Sigma_0$ is invertible, the minimization in Eq.~\eqref{eq:pipeau_final_stage} can be done analytically and gives $\widehat{A} = \Sigma_1 \Sigma_0^{-1}$. The estimate $\widehat{B}$ can then be obtained by solving \eqref{eq:max_likelihood_pipeau}.

\bibliographystyle{IEEEtran}
\bibliography{references.bib}{}

% Generated by IEEEtran.bst, version: 1.14 (2015/08/26)
\begin{thebibliography}{10}
\providecommand{\url}[1]{#1}
\csname url@samestyle\endcsname
\providecommand{\newblock}{\relax}
\providecommand{\bibinfo}[2]{#2}
\providecommand{\BIBentrySTDinterwordspacing}{\spaceskip=0pt\relax}
\providecommand{\BIBentryALTinterwordstretchfactor}{4}
\providecommand{\BIBentryALTinterwordspacing}{\spaceskip=\fontdimen2\font plus
\BIBentryALTinterwordstretchfactor\fontdimen3\font minus
  \fontdimen4\font\relax}
\providecommand{\BIBforeignlanguage}[2]{{%
\expandafter\ifx\csname l@#1\endcsname\relax
\typeout{** WARNING: IEEEtran.bst: No hyphenation pattern has been}%
\typeout{** loaded for the language `#1'. Using the pattern for}%
\typeout{** the default language instead.}%
\else
\language=\csname l@#1\endcsname
\fi
#2}}
\providecommand{\BIBdecl}{\relax}
\BIBdecl

\bibitem{kundur1994power}
P.~Kundur, N.~J. Balu, and M.~G. Lauby, \emph{Power system stability and
  control}.\hskip 1em plus 0.5em minus 0.4em\relax McGraw-hill New York, 1994,
  vol.~7.

\bibitem{sauer2017power}
P.~W. Sauer, M.~A. Pai, and J.~H. Chow, \emph{Power System Dynamics and
  Stability: With Synchrophasor Measurement and Power System Toolbox}.\hskip
  1em plus 0.5em minus 0.4em\relax John Wiley \& Sons, 2017.

\bibitem{huang2009application}
Z.~Huang, P.~Du, D.~Kosterev, and B.~Yang, ``Application of extended kalman
  filter techniques for dynamic model parameter calibration,'' in \emph{Power
  \& Energy Society General Meeting, 2009. PES'09. IEEE}.\hskip 1em plus 0.5em
  minus 0.4em\relax IEEE, 2009, pp. 1--8.

\bibitem{zhou2011calibration}
N.~Zhou, S.~Lu, R.~Singh, and M.~A. Elizondo, ``Calibration of reduced dynamic
  models of power systems using phasor measurement unit (pmu) data,'' in
  \emph{North American Power Symposium (NAPS), 2011}.\hskip 1em plus 0.5em
  minus 0.4em\relax IEEE, 2011, pp. 1--7.

\bibitem{guo2014adaptive}
S.~Guo, S.~Norris, and J.~Bialek, ``Adaptive parameter estimation of power
  system dynamic model using modal information,'' \emph{IEEE Transactions on
  Power Systems}, vol.~29, no.~6, pp. 2854--2861, 2014.

\bibitem{zhou2015dynamic}
N.~Zhou, D.~Meng, Z.~Huang, and G.~Welch, ``Dynamic state estimation of a
  synchronous machine using pmu data: A comparative study,'' \emph{IEEE
  Transactions on Smart Grid}, vol.~6, no.~1, pp. 450--460, 2015.

\bibitem{chen2016measurement}
Y.~C. Chen, J.~Wang, A.~D. Dom{\'\i}nguez-Garc{\'\i}a, and P.~W. Sauer,
  ``Measurement-based estimation of the power flow jacobian matrix,''
  \emph{IEEE Transactions on Smart Grid}, vol.~7, no.~5, pp. 2507--2515, 2016.

\bibitem{chavan2017identification}
G.~Chavan, M.~Weiss, A.~Chakrabortty, S.~Bhattacharya, A.~Salazar, and
  F.~Habibi-Ashrafi, ``Identification and predictive analysis of a multi-area
  wecc power system model using synchrophasors,'' \emph{IEEE Transactions on
  Smart Grid}, 2017.

\bibitem{wang2017pmu}
X.~Wang, J.~Bialek, and K.~Turitsyn, ``Pmu-based estimation of dynamic state
  jacobian matrix and dynamic system state matrix in ambient conditions,''
  \emph{IEEE Transactions on Power Systems}, 2017.

\bibitem{machowski1997power}
J.~Machowski, J.~Bialek, and J.~R. Bumby, \emph{Power system dynamics and
  stability}.\hskip 1em plus 0.5em minus 0.4em\relax John Wiley \& Sons, 1997.

\bibitem{chiang2011direct}
H.-D. Chiang, \emph{Direct methods for stability analysis of electric power
  systems: theoretical foundation, BCU methodologies, and applications}.\hskip
  1em plus 0.5em minus 0.4em\relax John Wiley \& Sons, 2011.

\bibitem{ghanavati2016identifying}
G.~Ghanavati, P.~D. Hines, and T.~I. Lakoba, ``Identifying useful statistical
  indicators of proximity to instability in stochastic power systems,''
  \emph{IEEE Transactions on Power Systems}, vol.~31, no.~2, pp. 1360--1368,
  2016.

\bibitem{van1998voltage}
T.~Van~Cutsem and C.~Vournas, \emph{Voltage stability of electric power
  systems}.\hskip 1em plus 0.5em minus 0.4em\relax Springer Science \& Business
  Media, 1998, vol. 441.

\bibitem{poolla2017optimal}
B.~K. Poolla, S.~Bolognani, and F.~Dorfler, ``Optimal placement of virtual
  inertia in power grids,'' \emph{IEEE Transactions on Automatic Control},
  2017.

\bibitem{deka2017acc}
D.~Deka, H.~Nagarajan, and S.~Backhaus, ``Optimal topology design for
  disturbance minimization in power grids,'' in \emph{American Control
  Conference (ACC)}, May 2017, pp. 2719--2724.

\bibitem{mendoza2016applying}
S.~Mendoza-Armenta and I.~Dobson, ``Applying a formula for generator redispatch
  to damp interarea oscillations using synchrophasors,'' \emph{IEEE
  Transactions on Power Systems}, vol.~31, no.~4, pp. 3119--3128, 2016.

\bibitem{chow2013power}
J.~H. Chow, \emph{Power system coherency and model reduction}.\hskip 1em plus
  0.5em minus 0.4em\relax Springer, 2013.

\bibitem{dorfler2013kron}
F.~Dorfler and F.~Bullo, ``Kron reduction of graphs with applications to
  electrical networks,'' \emph{IEEE Transactions on Circuits and Systems I:
  Regular Papers}, vol.~60, no.~1, pp. 150--163, 2013.

\bibitem{singh2010statistical}
R.~Singh, B.~C. Pal, and R.~A. Jabr, ``Statistical representation of
  distribution system loads using gaussian mixture model,'' \emph{IEEE
  Transactions on Power Systems}, vol.~25, no.~1, pp. 29--37, 2010.

\bibitem{deka2017state}
D.~Deka, A.~Zare, A.~Lokhov, M.~Jovanovic, and M.~Chertkov, ``State and noise
  covariance estimation in power grids using limited nodal pmus,'' in
  \emph{IEEE Global Conference on Signal and Information Processing}, 2017.

\bibitem{bento2010learning}
J.~Bento, M.~Ibrahimi, and A.~Montanari, ``Learning networks of stochastic
  differential equations,'' in \emph{Advances in Neural Information Processing
  Systems}, 2010, pp. 172--180.

\bibitem{Vauhkonen1998Tikhonov}
M.~Vauhkonen, D.~Vadasz, P.~A. Karjalainen, E.~Somersalo, and J.~P. Kaipio,
  ``Tikhonov regularization and prior information in electrical impedance
  tomography,'' \emph{IEEE Transactions on Medical Imaging}, vol.~17, no.~2,
  pp. 285--293, April 1998.

\bibitem{jalali2011learning}
A.~Jalali and S.~Sanghavi, ``Learning the dependence graph of time series with
  latent factors,'' in \emph{Proceedings of the 29th International Conference
  on Machine Learning (ICML)}, 2012, pp. 473--480.

\bibitem{demetriou2015dynamic}
P.~Demetriou, M.~Asprou, J.~Quiros-Tortos, and E.~Kyriakides, ``Dynamic ieee
  test systems for transient analysis,'' \emph{IEEE Systems Journal}, 2015.

\end{thebibliography}

%
% <OR> manually copy in the resultant .bbl file
% set second argument of \begin to the number of references
% (used to reserve space for the reference number labels box)

% that's all folks
\end{document}